\documentclass{IEEEtran}
\usepackage{cite}
\usepackage{amsmath,amssymb,amsfonts}
\usepackage{algorithmic}
\usepackage{graphicx}
\usepackage{textcomp}
\def\BibTeX{{\rm B\kern-.05em{\sc i\kern-.025em b}\kern-.08em
    T\kern-.1667em\lower.7ex\hbox{E}\kern-.125emX}}
\begin{document}
\title{Single-Photon Rydberg Excitation and Trap-Loss Spectroscopy of Cold Cesium Atoms in a Magneto-Optical Trap by Using of a 319-nm Ultra-Violet Laser System}
\author{Jiandong Bai, Shuo Liu, Jieying Wang, Jun He, and Junmin Wang*
\thanks{This work was supported in part by the National Key Research and Development Program of China under Grant 2017YFA0304502, in part by the National Natural Science Foundation of China under Grants 61475091, 11774210, and 61875111, and in part by the Shanxi Provincial 1331 Project for Key Subjects Construction.}
\thanks{Jiandong Bai, Shuo Liu, and Jieying Wang are with the State Key Laboratory of Quantum Optics and Quantum Optics Devices, and Institute of Opto-Electronics, Shanxi University, Tai Yuan 030006, China (e-mail: physjdbai@163.com; 1621488526@qq.com; wjy3861@163.com).}
\thanks{Jun He and Junmin Wang are with the State Key Laboratory of Quantum Optics and Quantum Optics Devices, and Institute of Opto-Electronics, Shanxi University, Tai Yuan 030006, China, and also with the Collaborative Innovation Center of Extreme Optics, Shanxi University, Tai Yuan 030006, China (e-mail: hejun@sxu.edu.cn; wwjjmm@sxu.edu.cn).}
\thanks{*Corresponding author: Junmin Wang (e-mail: wwjjmm@sxu.edu.cn).}}
\maketitle

\begin{abstract}
We demonstrate the single-photon Rydberg excitation of cesium atoms in a magneto-optical trap (MOT). We excite atoms directly from ${{6S}_{1/2}}$ ground state to ${{nP}_{3/2}}(n=70-100)$ Rydberg state with a narrow-linewidth 319 nm ultra-violet laser. The detection of Rydberg states is performed by monitoring the reduction of fluorescence signal of the MOT as partial population on ${{6S}_{1/2}} (F = 4)$ ground state are transferred to Rydberg state. We clearly observe Autler-Townes doublet in the trap-loss spectra due to the cooling lights. Utilizing the large electric polarizibility of Rydberg atoms, we observe Stark splitting in the Autler-Townes doublet induced by background DC electric fields. We investigate the dependence of Stark shift on electric fields by theoretical analysis, and then infer the DC electric field from the measured Stark splitting. We find that there is a 44.8(4) mV/cm DC electric field in the vicinity of the cold ensemble. It indicates that high-lying Rydberg atoms can be used as sensors for DC electric fields.
\end{abstract}

\begin{IEEEkeywords}
Single-photon Rydberg excitation, magneto-optical trap, trap-loss spectra, Stark effect, electric field sensing.
\end{IEEEkeywords}

\section{Introduction}
\label{sec:introduction}
\IEEEPARstart{R}{ydberg} atoms with principal quantum numbers $n>10$ have many exotic characters in terms of their unique structure. This includes large polarizability and microwave-transition dipole moments and lower field ionization threshold, which is proportional to ${{n}^{7}}$, ${{n}^{2}}$, and ${{n}^{-4}}$ [1], respectively, giving rise to an extreme sensitivity to the electric field environment. It makes Rydberg atoms promising for applications such as quantum information [2] and quantum metrology [3]-[6]. Due to the invariance of atomic properties, the atom-based field measurement has obvious advantages over other methods because it is calibration-free. Moreover, atom-based metrology has made significant progress in resolution, accuracy, and reproducibility, making it expanded into electric-field sensing. However, most experiments diagnosed DC electric field by the field ionization of high
Rydberg states [7]-[11]. While this detection method has high efficiency and discrimination, the detected atoms after field ionization has been destroyed and cannot be reused. Moreover, this detection system is more complex, and it is difficult to be miniaturization due to larger space usage. So it is necessary to perform the nondestructive detection of Rydberg states for the involved quantum information applications.

For purely optical detection of Rydberg states in a cold atomic sample, commonly employed scheme is the cascaded two-photon excitation [12], [13]. Two-photon excitation can avoid ultraviolet (UV) wavelengths making it easier to implement technically. However, it has following drawbacks: atomic decoherence due to the photon scattering from the lower transition via intermediate state. If we utilize off-resonant cascaded two-photon excitation to reduce the decoherence effect, it will introduce the light shift of the involved ground and Rydberg states, leading to a lower Rydberg excitation efficiency. Alternatively, direct excitation from the ground state to the desired highly-excited Rydberg state can avoid these drawbacks, but it demands narrow-linewidth tunable UV laser. Thanks to the development of the efficient laser frequency conversion technology and the fiber lasers as well as fiber amplifiers, high-power narrow-linewidth continuous-wave UV laser has been implemented [14]-[17]. Recently, Hankin $et$ $al.$ performed the single-photon Rydberg excitation using a continuous-wave UV laser, and demonstrated the coherent control of two single atoms in the ${{84P}_{3/2}}$ states [15], [18]. However, compared with the cold atomic ensemble, the signal-to-noise ratio of the spectrum based on the single atoms is lower.

In this paper, we demonstrate the single-photon Rydberg excitation of cesium cold atomic ensemble with a purely optical detection method, instead of ionization detection. We excite atoms directly from ${{6S}_{1/2}}$ ground state to  ${{nP}_{3/2}}$ ($n$ = 70-100) Rydberg state with a narrow-linewidth 319 nm UV laser. The detection of Rydberg states is performed by monitoring the fraction of remaining atoms in the ground state. Utilizing the trap-loss spectra technique, we clearly observe the Rydberg spectra of ${{nP}_{3/2}}$ states and predict the existence of background DC electric fields which will break degeneracies. Comparing the measured Stark splitting with the theoretical model, we can estimate the background DC electric field.

\begin{figure}[!t]
\centering\includegraphics[width=\columnwidth]{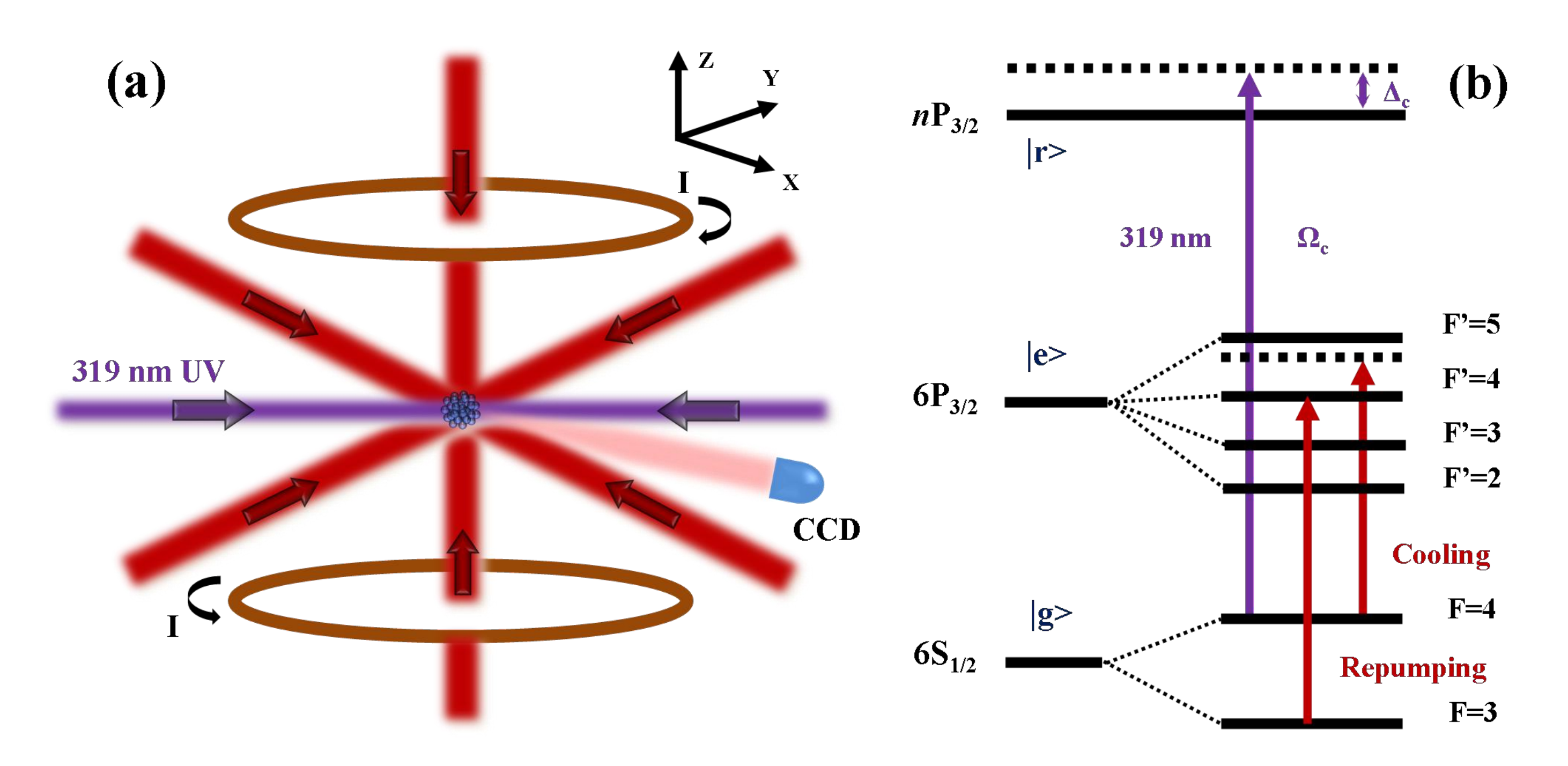}
\caption{(a) Diagram of single-photon Rydberg excitation in Cs MOT, which is constructed from six cooling beams at 852 nm, a repumping beam and a set of anti-Helmholtz coils. I is the direct current (DC) electric current of the anti-Helmholtz coils. A 319 nm UV laser beam is propagated through the MOT cloud, coupling the $\mid$$g\rangle$$\leftrightarrow\mid$$r\rangle$ transition. The fluorescence from the MOT is detected with a digital CCD camera. (b) Relative level scheme for Cs atomic single-photon Rydberg excitation. The 319 nm UV beam is linearly-polarized and near-resonant to the ${{6S}_{1/2}}(F = 4)\rightarrow{{nP}_{3/2}}$ transitions.}
\label{fig1}
\end{figure}

\section{Experimental arrangement}

We employ a cold Cs ensemble to perform the single-photon Rydberg excitation experiment. The experimental setup and the relative levels are schematically depicted in Fig. 1. The cold Cs atomic sample is prepared in a magneto-optical trap (MOT). The six cooling/trapping beams with total power of $\sim$120 mW and a typical beam diameter of $\sim$20 mm ($1/{{e}^{2}}$) are provided by a grating feedback external cavity diode laser (ECDL) (Toptica, DL100) which is red detuned by 12.4 MHz from Cs ${{6S}_{1/2}}(F = 4)\rightarrow{{6P}_{3/2}}(F' = 5)$ hyperfine transition. The repumping beam, overlapped with one of the cooling beams, is provided by a distributed-Bragg-reflector (DBR) diode laser which is stabilized to ${{6S}_{1/2}}(F = 3)\rightarrow{{6P}_{3/2}}(F' = 4)$ hyperfine transition by means of polarization spectroscopy [19]. The magnetic field gradient along axial direction of the anti-Helmholtz coils is 10 Gauss/cm in the vicinity of the atomic cloud. The trapped Cs atoms in the MOT have a Gaussian radius of $\sim$1.2 mm in a high-vacuum environment of $\sim$$2\times{{10}^{-9}}$ Torr. The density and number of cold atoms are on the order of $5\times{{10}^{10}}$ cm${{}^{-3}}$ and ${{10}^{9}}$. The temperature is measured to be $\sim$50 $\mu$K by a time-of-flight method. Excitation of Rydberg states is accomplished by a single-photon direct excitation scheme from ${{6S}_{1/2}}$ to ${{nP}_{3/2}}(n=70-100)$ state with a continuous-wave 319 nm UV beam, as shown in Fig. 1(b). The narrow-linewidth UV beam with a $1/{{e}^{2}}$ diameter of $\sim$2.4 mm is produced by the cavity-enhanced second-harmonic generation following sum-frequency generation of two infrared lasers at 1560 nm and 1077 nm, leading to more than 2 W output power at 319 nm [16]. To reduce the influence that the radiation pressure forces the atoms out of the trap, we use a 319 nm standing wave field to excite the atoms to Rydberg states.

Because Rydberg atoms cannot be trapped by the MOT [20], [21], Rydberg excitation will cause the atoms to be lost, resulting in the reduction of the atom number in ${{6S}_{1/2}}(F = 4)$ ground state. Therefore, we adopt high-precision trap-loss spectroscopy to determine the Rydberg excitation. The fluorescence intensity from the MOT is proportional to the population of ground-state atoms. By making the fluorescence measurement before and after the Rydberg excitation, the number of ground-state atoms lost can be measured. A digital CCD camera (Thorlabs, 1500M-GE) is used to monitor the atom number of the ground state [${{6S}_{1/2}}(F = 4)$] by taking spatially-resolved images of the MOT. A computer is exploited to control the camera parameters and record the data by analyzing the images' information collected by the CCD camera. During the execution of the experiment, we perform extensive tests to ensure the linearity of the camera response. While the MOT is continuously loaded by the cold atomic beams, it is exposed by a 319 nm UV beam. For each UV frequency, the MOT fluorescence is recorded on the CCD with and without UV beam for 10 s each. More details on the spectra will be discussed in the next section.

\section{Results and discussions}
\subsection{Trap-loss spectra of cesium MOT for single-photon Rydberg transition}

To study the Rydberg atomic characters and the interaction between them, it is of great importance to perform high-resolution spectroscopy. Generally, the continuous-scan and step-scan techniques are used to perform the spectroscopic measurement. The continuous-scan technique has some limitations: the unstable frequency will broaden the observed spectra, moreover, the continuous loading of the MOT during Rydberg excitation will result in complex population dynamics. This distorts the shape of the spectra. Compared to this technique, the step-scan technique has better signal-to-noise ratio and much narrower linewidth. The spectral lineshapes are highly symmetrical. Consequently, we employ the step-scan technique to perform Rydberg spectroscopy in the following way. In the process, the 319 nm UV beam is stabilized to a high-finesse ultra-low-expansion (ULE) optical cavity inside an ultra-high vacuum chamber using the electronic sideband locking method [22]. The relative frequency deviation of the UV beam is estimated to be less than 15 kHz. The high-stability narrow-linewidth UV beam can be continuously tuned over a range of 4 GHz while maintaining the lock.

\begin{figure}[!t]
\vspace{-0.05in}
\centering\includegraphics[width=\columnwidth]{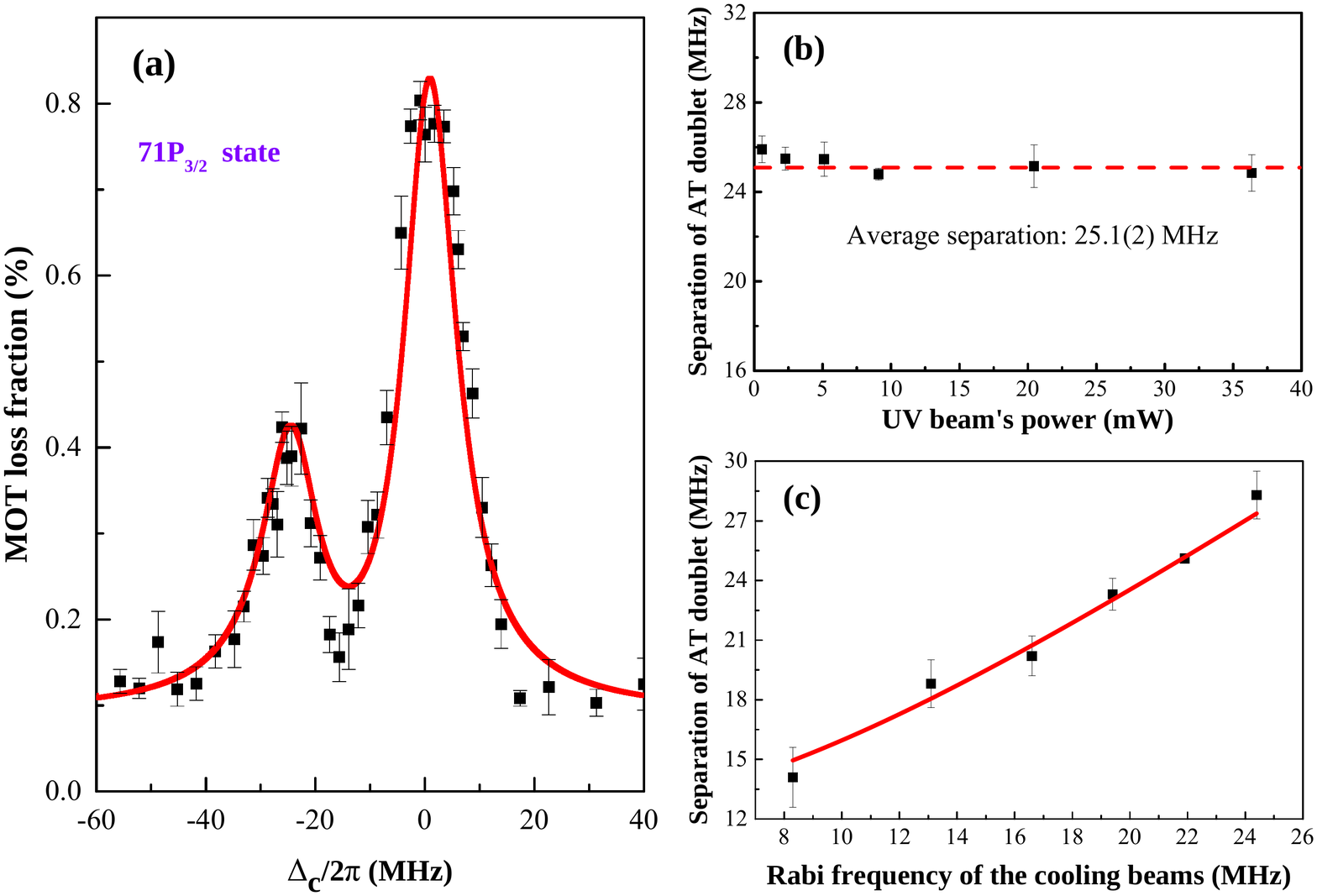}
\caption{(a) Trap-loss spectrum for cesium ${{6S}_{1/2}}(F = 4)\rightarrow{{71P}_{3/2}}$ single-photon Rydberg transition. The error bars are standard deviation via multi-measurement average. The experimental data are fitted by the theoretical calculation based on AT spectra in a V-type three-level system. The zero point of the relative UV frequency is chosen considering the light shift by the dressed state and DC Stark effect. (b) The separation of AT doublet for the ${{71P}_{3/2}}$ state versus the UV beam's power. Because Rabi frequency of the UV coupling beam is much less than that of the cooling beams, the separation is independent of the UV coupling intensity. The separation is measured to be 25.1(2) MHz. (c) The separation of AT doublet for the ${{71P}_{3/2}}$ state as a function of Rabi frequency of the cooling beams. The black squares are the experimental data, and the red solid line is the theoretical calculation.}
\label{fig2}
\end{figure}

Employing the frequency-stabilized tunable UV laser system, we perform the trap-loss spectrum of Rydberg state ${{71P}_{3/2}}$, as shown in Fig. 2(a). Here, the UV beam's power is fixed at 1 mW. Because the cooling beams' Rabi frequency is greater than natural linewidth of the ${{6P}_{3/2}}$  state, the existence of the strong cooling beams results in Autler-Townes (AT) splitting in the ground state [${{6S}_{1/2}}(F = 4)$] and the excited state [${{6P}_{3/2}}(F' = 5)$]. While a weak 319 nm UV beam is scanned across the transition of ${{6S}_{1/2}}(F = 4)\rightarrow{{71P}_{3/2}}$, the spectrum exhibits two loss peaks. To verify this, we measure the separation of AT doublet for ${{71P}_{3/2}}$ state as a function of the UV coupling intensity (0-40 mW), as shown in Fig. 2(b). Because Rabi frequency of the UV coupling beam is much less than that of the cooling beams, the separation of AT doublet is independent of the UV coupling intensity. Moreover, we also measure the separation of AT doublet at different Rabi frequencies of the cooling beams, as shown in Fig. 2(c). The calculation results (red solid line) by the dressed state theory [23] are consistent with the experimental data (black squares).
This further proves that the doublet is caused by the strong cooling beams which result in the dressed splitting of the ground state. The separation of AT doublet in Fig. 2(a) is measured to be approximately 25.1(2) MHz, which is in accord with the calculated value of 25.2 MHz. The splitting can be expressed as $\widetilde{\Omega} = \sqrt{{{\Omega}^{2}}+{{\Delta}^{2}}}$. Where $\Omega$ is the total Rabi frequency of the cooling beams, $\Delta$ is the detuning of cooling beams relative to ${{6S}_{1/2}}(F = 4)\rightarrow{{6P}_{3/2}}(F' = 5)$ hyperfine transition. The asymmetry of AT doublet is caused by the non-zero detuning of the cooling beams [23], [24]. In recent work [24], we also observed this phenomenon in a room-temperature vapor cell with moderate coupling intensity, and investigated the dependence of the separation and linewidth on the coupling intensity with dressed state theory. The experimental results showed good agreement with the theoretical prediction. Thus we don't elaborate it here.

To estimate the contribution from various broadening mechanisms, we fit the experimental data (black squares) in Fig. 2(a) by solving the density-matrix equations in a V-type three-level system which are discussed in detail in the ref. [23]. The spectral linewidth is about 14.7 MHz, including the natural
linewidth of 5.2 MHz from the ${{6P}_{3/2}}$ state, the UV beam's linewidth of $\sim$10 kHz [16], the Doppler width at the UV wavelength of ${{\Delta f}_{D}} = 2\sqrt{ln2} {{\upsilon}_{mp}}/\lambda\simeq$ 400 kHz, the Zeeman broadening of ${{6S}_{1/2}}$ state in the center of the MOT along the z direction (${{\Delta f}_{Z}} = \frac{{\mu}_{B}}{h} {{g}_{J}}({{S}_{1/2}}) \frac{{\partial B}_{z}}{\partial z} \Delta z\simeq$ 4.6 MHz with ${{g}_{J}}({{S}_{1/2}})$ = 2), and Stark broadening of 2.8 MHz from the background electric fields which will be discussed below. In the experiment, because the intensity of the 319-nm UV laser is weak, only some of the ground-state atoms in the MOT are excited to Rydberg state, resulting in a relatively low number density of Rydberg atoms. So the van der Waals interactions between Rydberg atoms have little effect on the spectral linewidth. Therefore, the other broadening of $\sim$1.7 MHz may contribute from the collisional broadening among the atoms.

\subsection{The influence of background DC electric field}

The orbital radius of Rydberg atoms scale to ${{n}^{2}}$ [1], and the spacings between adjacent states decrease sharply with increasing \emph{n}. This implies exaggerated sensitivity to DC electric field due to its large electric polarizability. We calculate that the electric polarizability of ${{71P}_{3/2}}$ state is on the order of ${{10}^{11}}$ times larger than that of ${{6S}_{1/2}}$ ground state. We exploit the large electric polarizability to assess the influence of the electric field environment in the vicinity of the MOT on the Rydberg spectra of ${{nP}_{3/2}}$ states.

We first model Stark splitting of Rydberg state by including the perturbing effects of electric field. The total Hamiltonian for an atom in a static electric field is written as:
\begin{equation}
H_{total}=H_{atom}+H_{Stark}
\end{equation}
Here, ${{H}_{atom}}$ is the Hamiltonian of bare atom and its matrix elements can be calculated using quantum defect theory [25]. The final term is the Stark interaction which is given by ${{H}_{Stark}}=-{{\mu}_{d}}E$. Here, ${{\mu}_{d}}$ and \emph{E} are the electric-dipole moment and a DC electric field, respectively. The eigenstates $\mid$$\chi\rangle$ of ${{H}_{total}}$ in a static electric field \emph{E} are a linear superposition of the unperturbed atomic $\mid$$n, s, l, j\rangle$ states:
\begin{equation}
\mid\chi(E)\rangle=\sum\limits_{n,l,j} c_{n,l,j}(E)\mid n, s, l, j\rangle
\end{equation}
Where ${{c}_{n,l,j}}$ are the mixing coefficients. The Stark interaction does not mix states of different total angular momentum projection number $\mid$${{m}_{j}}$$\mid$ and different total spin [26], so we can calculate a separate Stark map for different $\mid$${{m}_{j}}$$\mid$. The energy levels of different Rydberg states are much closer at higher \emph{n} (energy spacings scale as ${{n}^{-3}}$), so the Stark map is fairly complex in high electric fields. However, in low electric fields, Stark shift of the Rydberg state $\Delta\varepsilon$ can be given by second order perturbation theory as [27]:
\begin{equation}
\Delta\varepsilon=-\frac{1}{2}\alpha E^2
\end{equation}

The total polarizability {$\alpha$} is given by:
\begin{equation}
\alpha=\alpha_{0}+\alpha_{2}\frac{3{{m}_{j}}^{2}-j(j+1)}{j(2j-1)}
\end{equation}
Here, the energy shift is decided by a scalar polarizability $\alpha_{0}$, and the splitting of the energy level is given by a tensor polarizability $\alpha_{2}$ for different ${m}_{j}$ values. For ${nP}_{3/2}$ states, the total polarizabilities are expressed as $\alpha={{\alpha}_{0}}-{{\alpha}_{2}}$ for $\mid$${{m}_{j}}$$\mid$ = 1/2 and  $\alpha={{\alpha}_{0}}+{{\alpha}_{2}}$ for $\mid$${{m}_{j}}$$\mid$ = 3/2, respectively.

To accurately calculate the total polarizabilities, the scalar $\alpha_{0}$ and tensor $\alpha_{2}$ polarizabilities are evaluated as the sum over intermediate $\mid$$n', s', l', j'\rangle$ states allowed by the electric-dipole selection rules:
\begin{equation}
\alpha_{0}=\frac{2}{3(2j+1)}\sum_{n', l', j'}\frac{{\langle n', s', l', j'\parallel D\parallel n, s, l, j\rangle}^2}{E_{n', l', j'} - E_{n, l, j} }
\end{equation}

\begin{eqnarray}
\lefteqn{\alpha_{2}=\sqrt{\frac{40j(2j-1)}{3(j+1)(2j+1)(2j+3)}}
\sum_{n', l', j'} {(-1)}^{j+j'}}\nonumber\\ && \times{\begin{Bmatrix} j & 1 & j'\\ 1 & j & 2\end{Bmatrix}}
\frac{{\langle n', s', l', j'\parallel D\parallel n, s, l, j\rangle}^2}{E_{n', l', j'} - E_{n, l, j} }
\end{eqnarray}
Where $\langle n', s', l', j'\parallel D\parallel n, s, l, j\rangle$ is the electric-dipole matrix elements. \{...\} represents the Wigner six-j symbol. $E_{n, l, j}$ and $E_{n', l', j'}$ are zero-field energies of the $\mid$$n, s, l, j\rangle$ state and the dipole allowed intermediate $\mid$$n', s', l', j'\rangle$ state, respectively.

For the calculation of the energy shifts in the range of the given electric fields, we diagonalize (1) for a large enough set of states to ensure convergence of the eigenvectors and eigenvalues of ${{H}_{total}}$. We calculate states up to \emph{l} = 10, and a range of \emph{n} = $\pm$5. The simulated energy shifts for the $\mid$${{m}_{j}}$$\mid$ = 1/2, 3/2 components of cesium ${{nP}_{3/2}}$ state are shown in Fig. 3(a), which are consistent with the calculation results in the ref. [15], [28], [29]. Considering the contribution of tensor polarizability to the total polarizability, the ${nP}_{3/2}$ state is split into two components in an external electric field, corresponding to $\mid$${{m}_{j}}$$\mid$ = 1/2 and $\mid$${{m}_{j}}$$\mid$ = 3/2. Because the tensor polarizability is calculated to be negative, the $\mid$${{m}_{j}}$$\mid$ = 3/2 state has smaller polarizability than $\mid$${{m}_{j}}$$\mid$ = 1/2 from (4). That is, the Stark shift of $\mid$${{m}_{j}}$$\mid$ = 1/2 is langer than that of $\mid$${{m}_{j}}$$\mid$ = 3/2 in the same electric field. And the Stark shift becomes more severe with increasing \emph{n}. Moreover, the separation between the $\mid$${{m}_{j}}$$\mid$ = 1/2 and $\mid$${{m}_{j}}$$\mid$ = 3/2 components also increases with principal quantum number and DC electric fields [Fig. 3(b)].

\begin{figure}[!t]
\vspace{-0.25in}
\centering\includegraphics[width=\columnwidth]{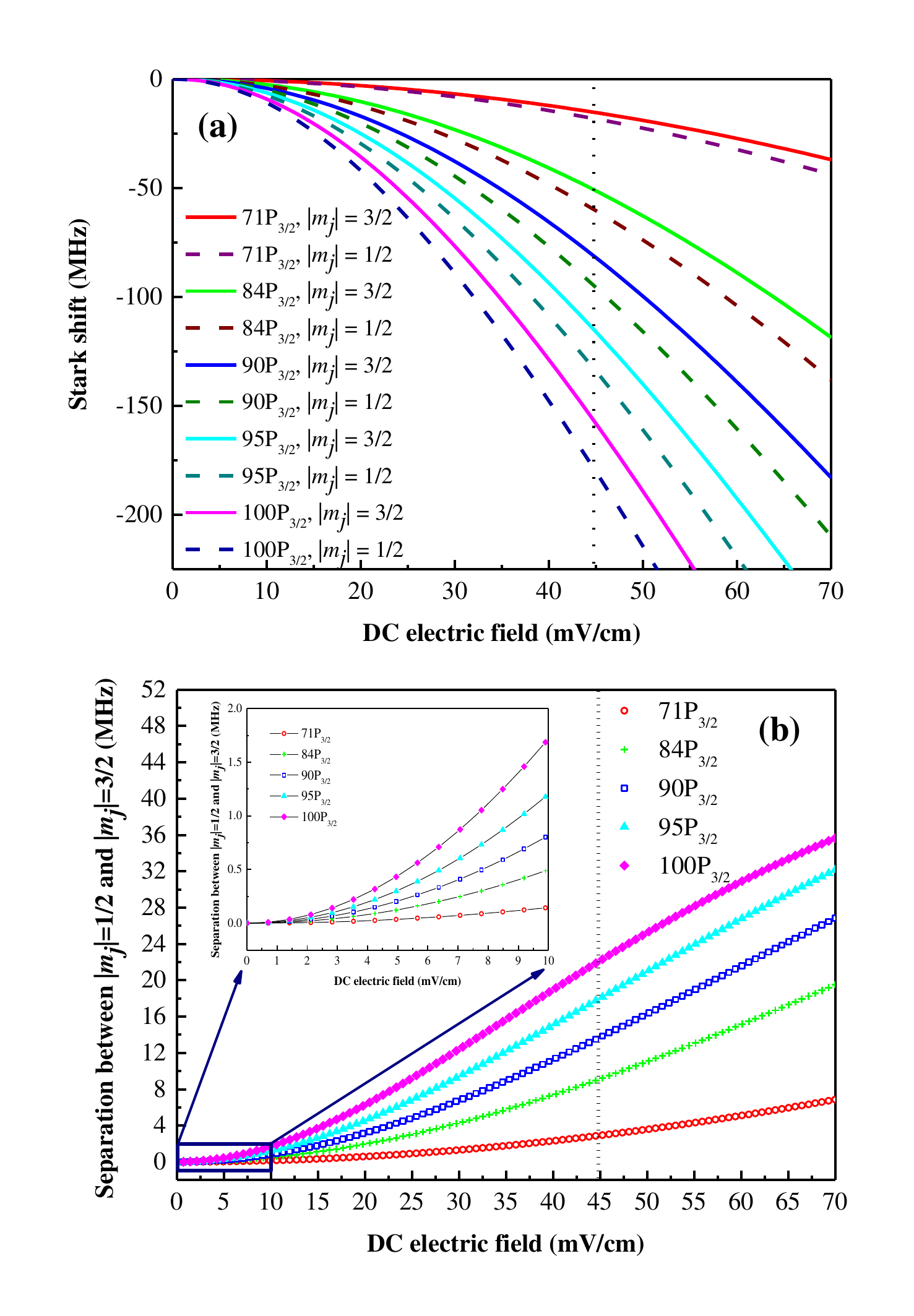}
\caption{(a) Calculated Stark shift for highly-excited cesium ${{nP}_{3/2}}$ (\emph{n}=71, 84, 90, 95 and 100) Rydberg states versus DC electric field. Energy shifts $\Delta\varepsilon$ are defined relative to the zero-field energy of each excited state. The solid lines indicate the energy shifts of the $\mid$${{m}_{j}}$$\mid$ = 3/2 component, while the dashed lines indicate that of the $\mid$${{m}_{j}}$$\mid$ = 1/2 component. (b) Separation between the $\mid$${{m}_{j}}$$\mid$ = 1/2 and $\mid$${{m}_{j}}$$\mid$ = 3/2 components with DC electric field. The illustration shows an enlarged view of DC electric fields between 0 and 10 mV/cm.}
\label{fig3}
\end{figure}

To determine the size of background DC electric field, we characterize the spectra of Rydberg states with different principal quantum numbers and compare the results with the theoretical prediction. The measured Rydberg spectra for ${{nP}_{3/2}}$ (\emph{n}=84, 90, 95 and 100) states are shown in Fig. 4. In order to determine the position of these peaks more accurately, a multi-peak Voigt function (red solid line) is exploited to fit the experimental data (black squares). We use a continuous-wave UV laser in a range of 100-MHz laser frequency to implement the Rydberg excitation experiment. Although we expect a single peak without any external perturbations, the observed spectrum is constituted of two nondegenerate sub-peaks (marked as 1, 2 or 3, 4) in any one of AT doublet. We use three sets of Helmholtz coils to compensate the geomagnetic field in the vicinity of the atomic cloud in three dimensions. Therefore, the degeneracy breaking of Rydberg spectra in Fig. 4 is due to the background DC electric field shifts the resonances by Stark effect.

Figure 4 shows that the spectrum shifts red and Stark splitting increases with \emph{n}. Because the separation of AT doublet is relatively small, the peak-3 gets closer to peak-2 with increasing \emph{n}, so that the two peaks gradually overlap in higher \emph{n}. Compared the measured Stark splitting (the spacings between the peak-1 and peak-2) with the theoretical value in Fig. 3(b), we can estimate the size of background DC electric field. The DC electric field in the vicinity of the atoms is 44.8 mV/cm with an uncertainty of $\pm$0.4 mV/cm, which is an average for five Rydberg states, as shown in Fig. 5. The error is mainly caused by the fluctuation of background DC electric field and the intensity fluctuation of the UV beam. The experimental data are in excellent agreement with the theory.

\begin{figure}[!t]
\vspace{-0.2in}
\centering\includegraphics[width=\columnwidth]{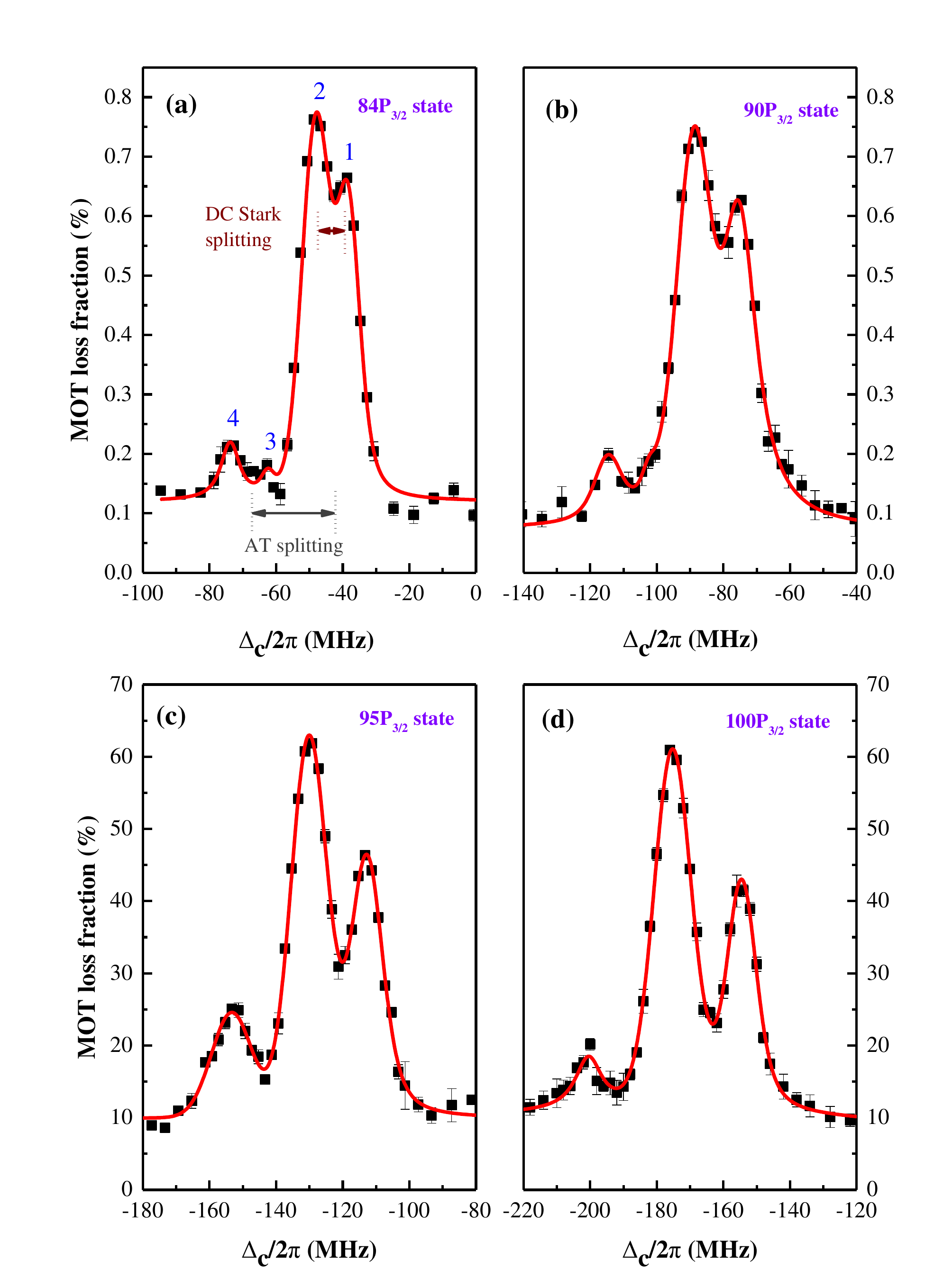}
\caption{The trap-loss spectra of cesium ${{nP}_{3/2}}$ (\emph{n} = 84, 90, 95 and 100) Rydberg states. The spectrum shifts red and Stark splitting increases with \emph{n}. Because the separation of AT doublet is relatively small, the peak-3 is much closer to peak-2 with increasing \emph{n}, so that the two peaks gradually overlap in higher principal quantum number (\emph{n} = 95 and 100).}
\label{fig4}
\end{figure}

\begin{figure}[!t]
\vspace{-0.2in}
\centering\includegraphics[width=\columnwidth]{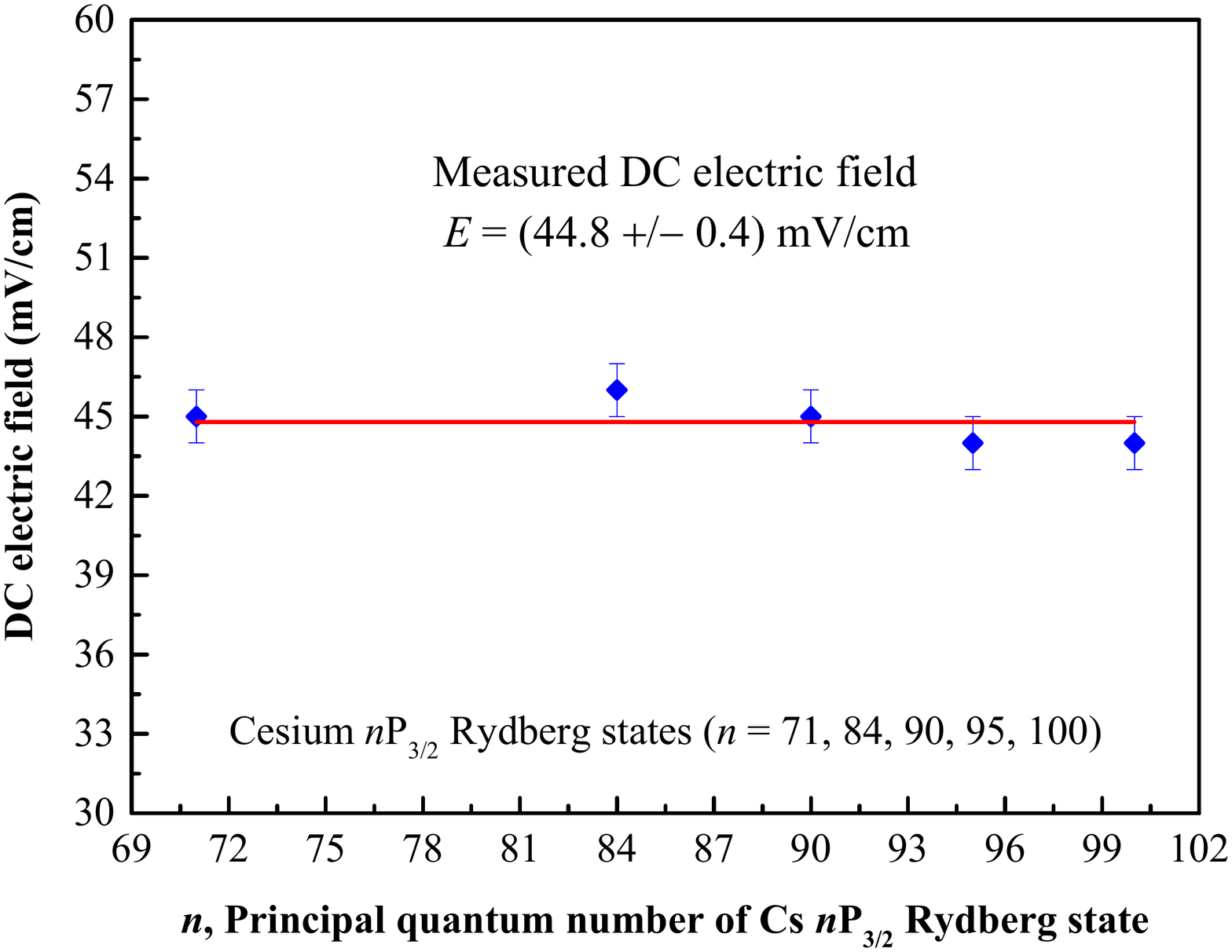}
\caption{Sensing the background DC electric field using highly-excited cesium ${{nP}_{3/2}}$ (\emph{n}=71, 84, 90, 95 and 100) Rydberg states. The red line represents the mean value for experimental data (blue diamond cube). From the measured Stark splitting, the background DC electric field is deduced to be 44.8 mV/cm, with an uncertainty of $\pm$0.4 mV/cm. The error is mainly due to the fluctuation of background DC electric field and the intensity fluctuation of the UV beam.}
\label{fig5}
\end{figure}

To characterize the sensitivity of the whole system to DC electric field, Rayleigh criterion is commonly used to define the minimum resolution of the spectrum. The criterion states that when the center of one peak falls on the edge of the other, the two peaks can just be distinguished. For ${{100P}_{3/2}}$ Rydberg state, the spectral linewidth is about 10.7 MHz for $\mid$${{m}_{j}}$$\mid$ = 1/2 or 3/2, leading to a minimum detectable field strength of $\sim$27 mV/cm. However, when the two peaks of Stark map overlap, the spectral linewidth will be squeezed with the reduced electric field. Thus, the measurement sensitivity of electric field is mainly limited to the frequency and intensity noises of the UV beam. We estimate that the sensitivity of DC electric field can be up to $\sim$1 mV/cm for ${{100P}_{3/2}}$ Rydberg state. For the smaller DC electric field measurement, Schramm \emph{et al.} [10] reduced electric field to $\sim$0.1 mV/cm by exciting argon atoms to \emph{n} = 400 Rydberg state. For \emph{n} = 400, its polarizability is on the order of ${{10}^{4}}$ times larger than that of \emph{n} = 100 Rydberg state. Therefore, we can also excite the atoms to higher Rydberg states to enhance the atomic sensitivity to DC electric fields. The measurement sensitivity of DC electric field will be better than $\sim$7 $\mu$V/cm for Cs Rydberg state with \emph{n} = 400. It indicates that Rydberg atoms have promising for applications in sensing unknown electric fields and can provide atom-based traceable standard for the electrometry.

To further enhance the sensitivity of electric field measurements, we will try our best to improve the signal-to-noise ratio of the spectra and suppress spectral broadening of Rydberg excitation signal which is essential for precision spectroscopy. The following measures can be taken to improve measurement accuracy: firstly, the ground-state atoms are excited by using of the short-pulsed UV light after turning off the gradient magnetic field and the trapping beams. In this way, it can not only eliminate the dressed splitting caused by the strong cooling lights, but also suppress the effect of Zeeman broadening on the spectra. Secondly, a fast feedback loop can be constructed to improve the UV frequency stability and narrow the laser linewidth [30]. In addition, we can use three pairs of electrode plates to compensate the background stray electric field in the vicinity of the atoms for future work. Then we apply a DC electric field to one pair of electrodes to accurately assess the minimum measurable value of electric field.

\section{Conclusion}

In summary, we have demonstrated the purely all-optical detection of cesium Rydberg atoms in a MOT via the single-photon excitation, instead of ionization detection. Utilizing a home-made narrow-linewidth 319 nm UV laser system, we excite cesium atoms directly from ${{6S}_{1/2}}$ ($F = 4$) ground state to ${{nP}_{3/2}}$ ($n$ = 70-100) Rydberg state. We clearly observe Autler-Townes splitting in trap-loss spectra due to the cooling beams. Moreover, we see the distinct Stark splitting of Rydberg spectra in high principal quantum number due to the influence of background DC electric field. We investigate the dependence of Stark shift on DC electric fields, and then infer the background DC electric field from the measured Stark splitting. We find that there is a 44.8(4) mV/cm DC electric field at the location of cold ensemble. Rydberg atoms have very large electric polarizibility, therefore DC electric field induced Stark splitting of Rydberg spectra can be employed to sense the unknown DC electric field, while cold Rydberg Cs atoms serve as a sensor. Consequently, the precision spectroscopy of high-lying Rydberg atoms plays an important role in quantum metrology. Moderate improvements in the system and excitation of higher Rydberg states are expected to enable more precise DC electric field measurement.

\end{document}